%% file: article_full.tex
\documentclass[fleqn,12pt,twoside]{article}
\usepackage[headings]{espcrc1}
\usepackage{graphicx}

\newcommand{\bn}{\begin{displaymath}}            
\newcommand{\en}{\end{displaymath}}
\newcommand{\bq}{\begin{equation}}               
\newcommand{\eq}{\end{equation}}

\newcommand{\lbo}{\linebreak[1]}
\newcommand{\lbt}{\linebreak[2]}

\newcommand{\jostle}{{\sc Jostle}}               
\newcommand{\metis}{\mbox{\sc Me$\!$T$\!$iS}}    
\newcommand{\parmetis}{\mbox{\sc ParMe$\!$T$\!$iS}}
\newcommand{\pastix}{{\sc PaStiX}}               
\newcommand{\scotch}{{\sc Scotch}}               
\newcommand{\ptscotch}{{\sc PT-Scotch}}          
\newcommand{\scalapplix}{{\sc ScAlApplix}} 

\title{\ptscotch: A tool for efficient parallel graph ordering}

\author{C. Chevalier\address[LaBRI]{LaBRI \& Project ScAlApplix of INRIA Futurs\\
        351, cours de la Lib\'eration, 33400 Talence, France}%
        \thanks{This author's work is funded by a joint PhD grant of CNRS and R\'egion Aquitaine.},
        F. Pellegrini\address[ENSEIRB]{ENSEIRB, LaBRI \& Project ScAlApplix of INRIA Futurs\\
        351, cours de la Lib\'eration, 33400 Talence, France\\
        {\tt \{cchevali|pelegrin\}@labri.fr}}}
       
\begin{document}

\maketitle

\begin{abstract}
The parallel ordering of large graphs is a difficult problem, because
on the one hand minimum degree algorithms do not parallelize well, and
on the other hand the obtainment of high quality orderings with the nested
dissection algorithm requires efficient graph bipartitioning heuristics,
the best sequential implementations of which are also hard to
parallelize.
This paper presents a set of algorithms, implemented in the \ptscotch\
software package, which allows one to order large graphs in parallel,
yielding orderings the quality of which is only slightly worse than
the one of state-of-the-art sequential algorithms. Our implementation
uses the classical nested dissection approach but relies on several
novel features to solve the parallel graph bipartitioning
problem. Thanks to these improvements, \ptscotch\ produces
consistently better orderings than \parmetis\ on large numbers of
processors.
\end{abstract}




\section{Introduction}

Graph partitioning is an ubiquitous technique which has applications
in many fields of computer science and engineering. It is mostly used
to help solving domain-dependent optimization problems modeled in
terms of weighted or unweighted graphs, where finding good solutions
amounts to computing, eventually recursively in a divide-and-conquer
framework, small vertex or edge cuts that balance evenly the weights
of the graph parts.

Because there always exist large problem graphs which cannot fit in
the memory of sequential computers and cost too much to partition,
parallel graph partitioning tools have been
developed~\cite{webmetis-parmetis,webjostle}.
The purpose of the \ptscotch{} software (``{\it Parallel Threaded
\scotch}'', an extension of the sequential \scotch{}
software~\cite{webscotch}), developed at LaBRI within the \scalapplix\
project of INRIA Futurs, is to provide efficient parallel tools to
partition graphs with sizes up to a billion vertices, distributed over
a thousand processors. Because of this deliberately ambitious goal,
scalability issues have to receive much attention.

One of the target applications of \ptscotch\ within the
\scalapplix\ project is graph ordering, which is a critical problem
for the efficient factorization of symmetric sparse matrices, not only to
reduce fill-in and factorization cost, but also to increase
concurrency in the elimination tree, which is essential in order to
achieve high performance when solving these linear systems on parallel
architectures. We therefore focus in this paper on this specific problem,
although we expect some of the algorithms presented here to
be reused in the near future in the context of edge, k-way
partitioning.

The two most classically-used reordering methods are minimum degree
and nested dissection. The minimum degree algorithm~\cite{tiwa67}
is a local heuristic which is extremely fast and very
often efficient thanks to the many improvements which have been
brought to it~\cite{amdadu96,geli89,liu-85}, but it is intrinsically
sequential, so that attempts to derive parallel versions of it have
not been successful~\cite{chgito99}, especially for distributed-memory
architectures. The nested dissection method~\cite{geli81}, on the
contrary, is very suitable for parallelization, since it consists in
computing a small vertex set that separates the graph into two parts,
ordering the separator vertices with the highest indices available,
then proceeding recursively on the two separated subgraphs until their
size is smaller than a specified threshold.

This paper presents the algorithms which have been implemented in
\ptscotch{} to parallelize the nested dissection method. The
distributed data structures used by \ptscotch{} will be presented in
the next section, while the algorithms that operate on them will be
described in Section~\ref{secalgo}. Section~\ref{secresult} will show
some results, and the concluding section will be devoted to
discussing some on-going and future work.


\section{Distributed data structures}
\label{secstruct}

Since \ptscotch\ extends the graph ordering capabilities of \scotch\
in the parallel domain, it has been necessary to define parallel data
structures to represent distributed graphs as well as distributed
orderings.

\subsection{Distributed graph}

\begin{figure}
\includegraphics[scale=0.5]{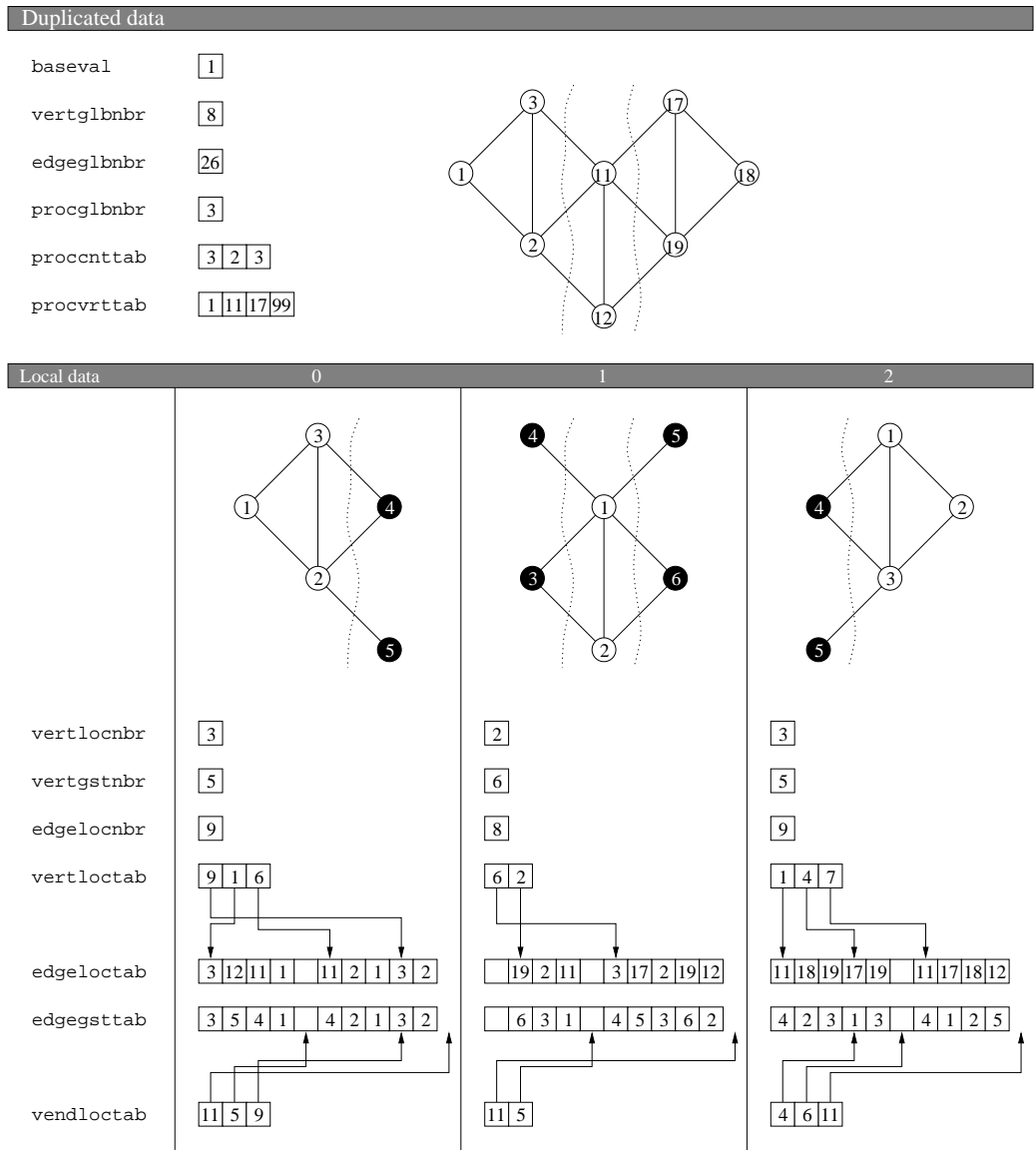} 
\\*[-1em]
\caption{Data structures of a graph distributed across three
  processes. The global image of the graph is shown above, while the
  three partial subgraphs owned by the three processes are
  represented below.
  Adjacency arrays with global vertex indexes are stored in
  {\tt edge\protect\lbt loc\protect\lbt tab} arrays, while local
  compact numberings of local and ghost neighbor vertices are
  internally available in {\tt edge\protect\lbt gst\protect\lbt tab}
  arrays. Local vertices owned by every process are drawn in
  white, while ghost vertices are drawn in black. For each local
  vertex $i$ located on process $p$, the global index of which is
  $({\tt proc\protect\lbt vrt\protect\lbt tab[}p{\tt ]} + i - {\tt
  base\protect\lbt val})$, the starting index of the adjacency array
  of $i$ in {\tt edge\protect\lbt loc\protect\lbt tab} (global
  indices) or {\tt edge\protect\lbt gst\protect\lbt tab} (local
  indices) is given by ${\tt vert\protect\lbt loc\protect\lbt
  tab[}i{\tt ]}$, and its after-end index by ${\tt vend\protect\lbt
  loc\protect\lbt tab[}i{\tt ]}$. For instance, local vertex $2$ on
  process $1$ is global vertex $12$; its start index in the
  adjacency arrays is $2$ and its after-end index is $5$; it has
  therefore $3$ neighbors, the global indices of which are $19$, $2$
  and $11$ in  {\tt edge\protect\lbt loc\protect\lbt tab}.}
\label{fig-struct-graph}
\end{figure}

Like for centralized graphs in \scotch\ as well as in other software
packages such as \metis~\cite{webmetis-parmetis}, distributed graphs
are classically represented in \ptscotch\ by means of adjacency
lists. Vertices are distributed across processes along with their
adjacency lists and with some duplicated global data, as illustrated
in Figure~\ref{fig-struct-graph}. In order to allow users to create
and destroy vertices without needing any global renumbering, every
process is assigned a user-defined range of global vertex
indices. Range arrays are duplicated across all processes in order to
allow each of them to determine the owner process of any non-local
vertex by dichotomy search, whenever necessary.

Since many algorithms require that local data be attached to every
vertex, and since global indices cannot be used for that purpose, all
vertices owned by some process are also assigned local indices,
suitable for the indexing of compact local data arrays. This local
indexing is extended so as to encompass all non-local vertices which
are neighbors of local vertices, which are referred to as ``ghost'' or
``halo'' vertices. Ghost vertices are numbered by ascending process
number and by ascending global number, such that, when vertex data
have to be exchanged between neighboring processes, these data can be
agglomerated in a cache-friendly way on the sending side, by
sequential in-order traversal of the data array, and be
received in place in the ghost data arrays on the receiving side.

A low-level halo exchange routine is provided by \ptscotch, to diffuse
data borne by local vertices to the ghost copies possessed by all of
their neighboring processes. This low-level routine is used by many
algorithms of \ptscotch, for instance to spread vertex labels of
selected vertices in the induced subgraph building routine (see
Section~\ref{secalgond}), or to share matching data in the coarse graph
building routine (see Section~\ref{secalgocoarsen}).

Because global and local indexings coexist, two adjacency arrays are
in fact maintained on every process. The first one, usually provided
by the user, holds the global indices of the neighbors of any given
vertex, while the second one, which is internally maintained by \ptscotch,
holds the local and ghost indices of the neighbors.
Since only local vertices are processed by the distributed algorithms,
the adjacency of ghost vertices is never stored on the processes,
which guarantees the scalability of the data structure as no process
will store information of a size larger than its number of local
outgoing arcs.

\subsection{Distributed ordering}

During its execution, \ptscotch\ builds a distributed tree structure,
spreading on all of the processes onto which it is run, and the
leaves of which represent fragments of the inverse permutation vector
describing the computed ordering. We use inverse permutation vectors
to represent orderings, rather than direct permutations, because
inverse permutations can be built and stored in a fully distributed
way.

Every subgraph to be reordered at some stage of the nested dissection
process is only provided with the global start index, in the inverse
permutation array, of the sub-ordering to compute on its vertices. If
the subgraph is too small to be separated or resides on a single
process, it is reordered using sequential methods, and the
corresponding inverse permutation fragment is built, of a size equal
to the number of vertices in the subgraph, and filled with the
original global indices of the reordered subgraph vertices, in local
inverse permutation order.
At the end of the nested dissection process, the assembly of all of
these fragments, by ascending start indices, yields the complete
inverse permutation vector.


\section{Algorithms for efficient parallel reordering}
\label{secalgo}

The parallel computation of orderings in \ptscotch\ involves three
different levels of concurrency, corresponding to three key steps of
the nested dissection process: the nested dissection algorithm itself,
the multi-level coarsening algorithm used to compute separators at
each step of the nested dissection process, and the refinement of the
obtained separators. Each of these steps is described below.

\subsection{Nested dissection}
\label{secalgond}

\begin{figure}
~\hfill%
\includegraphics[scale=0.44]{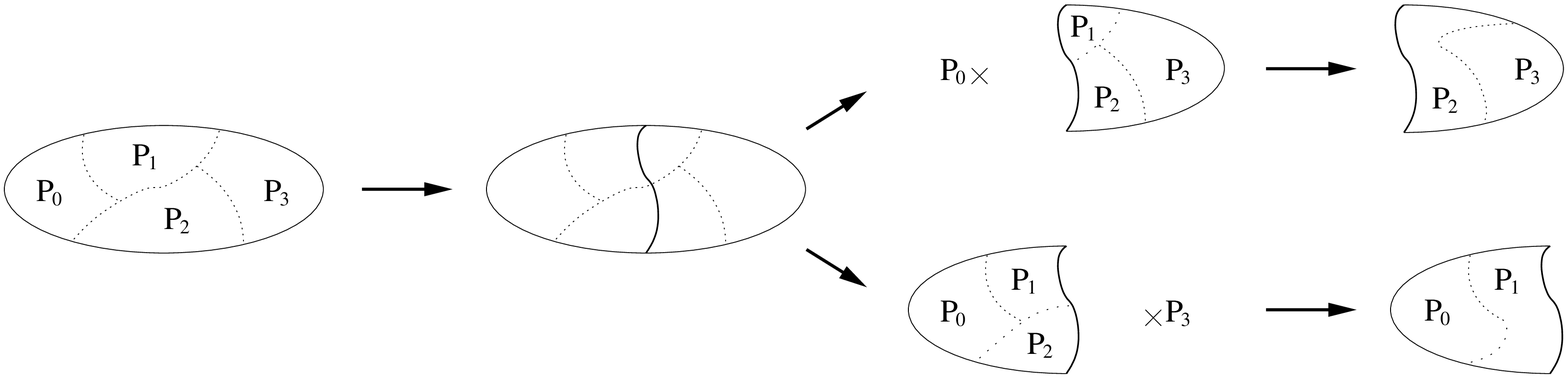}
\hfill~\\*[-1em]
\caption{Diagram of a nested dissection step for a (sub-)graph
  distributed across four processes.}
\label{fig-nedi}
\end{figure}

As said above, the first level of concurrency relates to the
parallelization of the nested dissection method itself, which is
straightforward thanks to the intrinsically concurrent nature of the
algorithm. Starting from the initial graph, arbitrarily distributed
across $p$ processes but preferably balanced in terms of vertices,
the algorithm proceeds as illustrated in Figure~\ref{fig-nedi}: once
a separator has been computed in parallel, by means of a method described
below, each of the $p$ processes participates in the building of the
distributed induced subgraph corresponding to the first separated part
(even if some processes do not have any vertex of it). This induced
subgraph is then folded onto the first $\lceil\frac{p}{2}\rceil$
processes, such that the average number of vertices per process,
which guarantees efficiency as it allows to overlap communications by
a subsequent amount of computation, remains constant. During the
folding phase, vertices and adjacency lists owned by the
$\lfloor\frac{p}{2}\rfloor$ sender processes are redistributed across
the $\lceil\frac{p}{2}\rceil$ receiver processes so as to evenly
balance their loads.

The same procedure is used to build, on the
$\lfloor\frac{p}{2}\rfloor$ remaining processes, the folded induced
subgraph corresponding to the second part. These two constructions
being completely independent, the computations of the two induced
subgraphs and their folding can be performed in parallel, thanks to the
temporary creation of an extra thread per process. When the vertices
of the separated graph are evenly distributed across the processes,
this feature favors load balancing in the subgraph building phase,
because processes which do not have many of their vertices in one part
have the rest of them in the other part, thus necessitating the
same overall amount of work to create both graphs in the same time. This
feature can be disabled when the communication system of the target
machine is not thread-safe.

At the end of the folding phase, every process has a
subgraph fragment of one of the two folded subgraphs, and the nested
dissection algorithm can recursively proceed independently on each
subgroup of $\frac{p}{2}$ (then $\frac{p}{4}$, $\frac{p}{8}$,
etc\@.) processes, until each subgroup is reduced to a single
process. From then on, the nested dissection algorithm will go on
sequentially on every process, using the nested dissection routines
of the \scotch\ library, eventually ending in a coupling with minimum
degree methods~\cite{peroam00a} (which are thus only used in a
sequential context).

\subsection{Graph coarsening}
\label{secalgocoarsen}

The second level of concurrency concerns the computation of
separators. The approach we have chosen is the now classical
multi-level one~\cite{basi94,hele95,kaku98a}. It consists in
repeatedly computing a set of increasingly coarser albeit
topologically similar versions of the graph to separate, by finding
matchings which collapse vertices and edges, until the coarsest graph
obtained is no larger than a few hundreds of vertices, then computing
a vertex separator on this coarsest graph, and projecting back this
separator, from coarser to finer graphs, up to the original graph.
Most often, a local optimization algorithm, such as
Kernighan-Lin~\cite{keli70} or Fiduccia-Mattheyses~\cite{fima82} (FM),
is used in the uncoarsening phase to refine the partition which is
projected back at every level, such that the granularity of the
solution is the one of the original graph and not the one of the
coarsest graph. Because we are interested in vertex separators, the
refinement algorithm we use is a vertex-oriented variant of the FM
algorithm, similar to the one described in~\cite{hero98}.

\begin{figure}
~\hfill%
\includegraphics[scale=0.35]{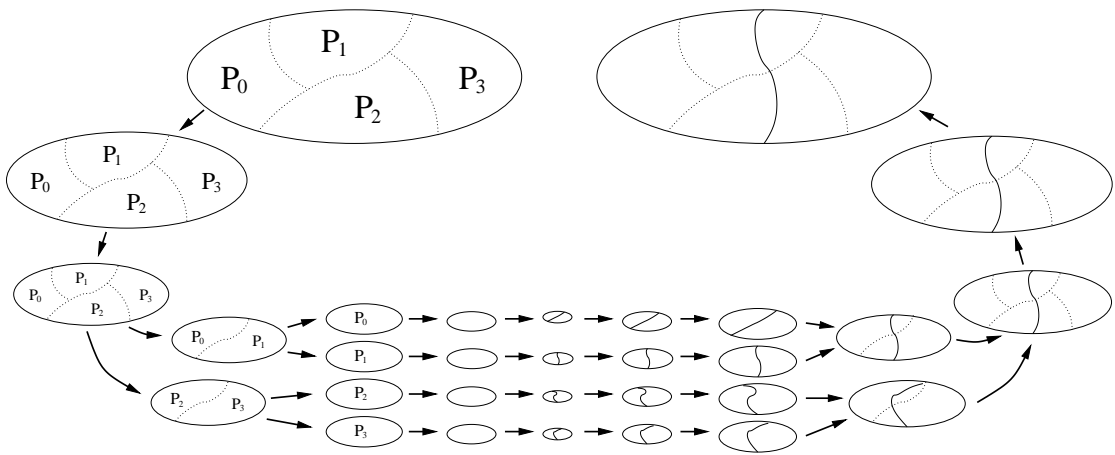}
\hfill~\\*[-1em]
\caption{Diagram of the parallel computation of the separator of a
  graph distributed across four processes, by parallel coarsening
  with folding-with-duplication, multi-sequential computation of
  initial partitions that are locally projected back and refined on
  every process, and then parallel uncoarsening of the best
  partition encountered.}
\label{fig-sepa}
\end{figure}

The main features of our implementation are outlined in
Figure~\ref{fig-sepa}. The matching of vertices is performed in
parallel by means of a synchronous probabilistic algorithm. Every
process works on a queue storing the yet unmatched vertices which it
owns, and repeats the following steps. The queue head vertex is
dequeued, and a candidate for mating is chosen among its unmatched
neighbors, if any; else, the vertex is left unmatched at this level
and discarded. When there are several unmatched neighbors, the
candidate is randomly chosen among vertices linked by edges of
heaviest weight, as in~\cite{kaku95a}. If the candidate vertex belongs
to the same process, the mating is immediatly recorded, else a
mating request is stored in a query buffer to be sent to the proper
neighbor process, and both vertices (the local vertex and its ghost
neighbor) are flagged as temporarily unavailable. Once all vertices in
queue have been considered, query buffers are exchanged between
neighboring processes, and received query buffers are processed in
order to satisfy feasible pending matings. Then, unsatisfied mating
requests are notified to their originating processes, which unlock
and reenqueue the unmatched vertices. This whole process is repeated
until the list is almost empty; we do not wait until it is completely
empty because it might require too many collective steps for just a
few remaining vertices. It usually converges in $5$ iterations.

The coarsening phase starts once the matching phase has converged. It
can be para\-me\-triz\-ed so as to allow one to choose between two
options. Either all coarsened vertices are kept on their local
processes (that is, processes that hold at least one of the ends of
the coarsened edges), as shown in the first steps of
Figure~\ref{fig-sepa}, which decreases the number of vertices owned by
every process and speeds-up future computations, or else coarsened
graphs are folded and duplicated, as shown in the next steps of
Figure~\ref{fig-sepa}, which increases the number of working copies of
the graph and can thus reduce communication and improve the final
quality of the separators. A folding algorithm is also implemented in
\parmetis, in the end of its multi-level process, to keep the number
of vertices per process sufficiently high so as to prevent running
time from being dominated by communications~\cite{kaku97}. However, in
its case, the folded subgraphs are not duplicated on the two subsets
of processes, and its folding algorithm requires the number of
sending processes to be even, such that the parallel graph ordering
routine of \parmetis\ can only work on a numbers of processes which
are powers of two. \ptscotch\ does not have this limitation, and can
run on any number of processes.

As a matter of fact, separator computation algorithms, which are local
heuristics, heavily depend on the quality of the coarsened graphs,
and we have observed with the sequential version of \scotch\ that
taking every time the best partition among two ones, obtained from
two fully independent multi-level runs, usually improves overall ordering
quality. By enabling the folding-with-duplication routine (which will
be referred to as ``fold-dup'' in the following) in the first
coarsening levels, one can implement this approach in parallel, every
subgroup of processes that hold a working copy of the graph being
able to perform an almost-complete independent multi-level
computation, save for the very first level which is shared by all
subgroups, for the second one which is shared by half of the subgroups,
and so on.

The problem with the fold-dup approach is that it
consumes a lot of memory. When no folding occurs, and in the ideal
case of a perfect and evenly balanced matching, the coarsening process
yields on every process a part of the coarser graph which is half
the size of the finer graph, and so on, such that the overall memory
footprint on every process is about twice the size of the original
graph. When folding occurs, every process receives two coarsened
parts, one of which belonging to another process, such that the size
of the folded part is about the one of the finer graph. The footprint
of the fold-dup scheme is therefore logarithmic in the
number of processes, and may consume all available memory as this
number increases. Consequently, as in~\cite{ParMetis}, a good strategy
can be to resort to folding only when the number of vertices of the
graph to be considered reaches some minimum threshold. This threshold
allows one to set a trade-off between the level of completeness of the
independent multi-level runs which result from the early stages of the
fold-dup process, which impact partitioning quality,
and the amount of memory to be used in the process.

Once all working copies of the coarsened graphs are folded on
individual processes, the algorithm enters a multi-sequential phase,
illustrated at the bottom of Figure~\ref{fig-sepa}: the routines of
the sequential \scotch\ library are used on every process to
complete the coarsening process, compute an initial partition, and
project it back up to the largest centralized coarsened graph stored
on each of the processes. Then, the partitions are projected back in
parallel to the finer distributed graphs, selecting the best partition
between the two available when projecting to a level where fold-dup
had been performed. This distributed projection process is repeated
until we obtain a partition of the original graph.

\subsection{Band refinement}

The third level of concurrency concerns the refinement heuristics
which are used to improve the projected separators. At the coarsest
levels of the multi-level algorithm, when computations are restricted to
individual processes, the sequential vertex FM algorithm of \scotch\ is
used, but this class of algorithms does not parallelize well. Indeed,
a parallel FM-like algorithm has been proposed in
\parmetis~\cite{kaku96} but, in order to relax the strong
sequential constraint that would require some communication every time
a vertex to be migrated has neighbors on other processes, only moves
that strictly improve the partition are allowed, which hinders the
ability of the FM algorithm to escape from local minima of its cost
function, and leads to severe loss of partition quality when the
number of processes (and thus of potential remote neighbors)
increases.

This problem can be solved in two ways: either by developing
scalable and efficient local optimization algorithms, or by being able
to use the existing sequential FM algorithm on very large graphs. We
have proposed and successfully tested in~\cite{chpe06a} a solution
which enables both approaches, and is based on the following
reasoning. Since every refinement is performed by means of a local
algorithm, which perturbs only in a limited way the position of the
projected separator, local refinement algorithms need only be
passed a subgraph that contains the vertices that are very close to
the projected separator. We have therefore implemented a distributed
band graph extraction algorithm, which only keeps vertices that are at
small distance from the projected separators, such that our local
optimization algorithms are applied to these band graphs rather than
to the whole graphs.

Using FM algorithms on band graphs substantially differs from what is
called ``boundary FM'' in the literature~\cite{hele93c,kaku98b}. This
latter technique amounts in FM-like algorithms to recomputing gains of
vertices which are in the immediate vicinity of the current separator
only, in order to save computation time, and \scotch\ also benefits
from this optimization, even on band graphs themselves. What differs
with our use of band graphs is that feasible moves are limited to the
band area, from which refined separators will never move away, while
they could move far away from projected separators in the case of
unconstrained FM, whether boundary-optimized or not.

We have experimented that, when performing FM refinement on band
graphs that contain vertices that are at distance at most $3$ from the
projected separators, the quality of the finest separator does not
only remain constant, but even improves in most cases, sometimes
significantly. Our interpretation is that pre-constrained banding
prevents local optimization algorithms from exploring and being
trapped in local optima that would be too far from the global optimum
sketched at the coarsest level of the multi-level process. The
optimal band width value of $3$ that we have evidenced is significant
in this respect: it is the maximum distance at which two vertices can
be in some graph when the coarse vertices to which they belong are
neighbors in the coarser graph of next level. Therefore, keeping more
layers of vertices in the band graph is not useful, because allowing
the fine separator to move at a distance greater than $3$ in the fine
graph to reach some local optimum means that it could already have moved
to this local optimum in the coarser graph, save for coarsening
artefacts, which are indeed what we want not to be influenced by.
\\

The advantage of pre-constrained band FM is that band graphs are of a
much smaller size than their parent graphs, since for most graphs the
size of the separators is of several orders of magnitude smaller that
the size of the separated graphs: it is for instance in
$O(n^\frac{1}{2})$ for 2D meshes, and in $O(n^\frac{2}{3})$ for 3D
meshes~\cite{lirota79}. Consequently, FM or other algorithms can be
run on graphs that are much smaller, without decreasing separation
quality.

\begin{figure}
~\hfill%
\includegraphics[scale=0.44]{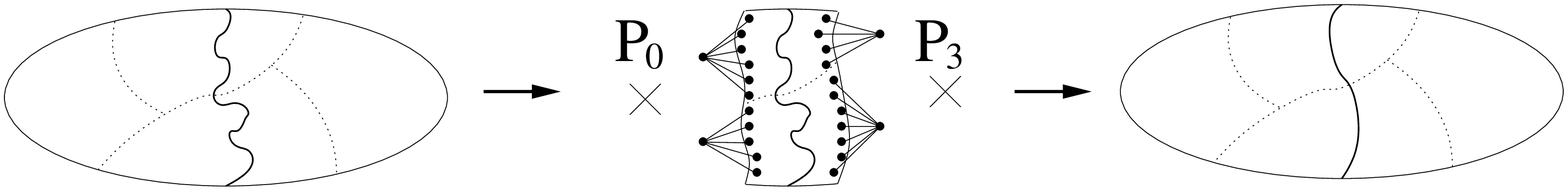}
\hfill~\\*[-1em]
\caption{Creation of a distributed band graph from a projected
  partition. Once the separator of the band graph is refined, it
  is projected back to the original distributed graph.}
\label{fig-band}
\end{figure}

The computation and use of distributed band graphs is outlined in
Figure~\ref{fig-band}. Given a distributed graph and a projected
separator, which can be spread across several processes, vertices
that are closer to separator vertices than some small user-defined
distance are selected by spreading distance information from all of
the separator vertices, using our halo exchange routine. Then, the
distributed band graph is created, by adding on every process two
anchor vertices, which are connected to the last layers of vertices of
each of the parts. The vertex weight of the anchor vertices is equal
to the sum of the vertex weights of all of the vertices they replace,
to preserve the balance of the two band parts. Once the separator of
the band graph has been refined using some local optimization
algorithm, the new separator is projected back to the original
distributed graph.

\begin{figure}
~\hfill%
\includegraphics[scale=0.30]{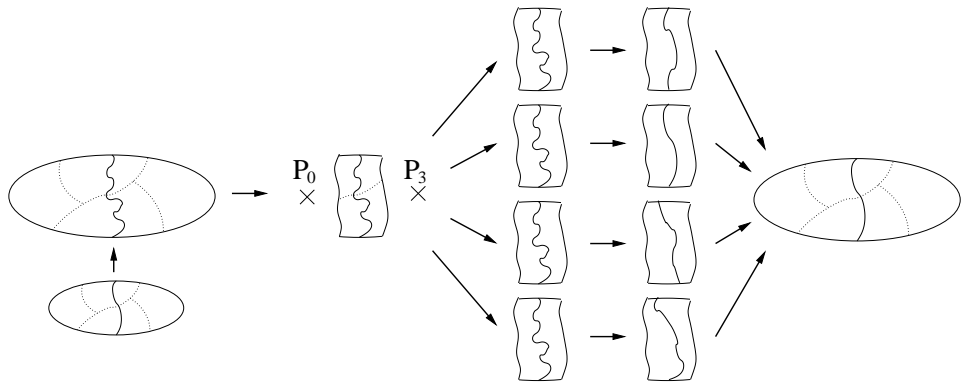}
\hfill~\\*[-1em]
\caption{Diagram of the multi-sequential refinement of a separator
  projected back from a coarser graph distributed across four processes
  to its finer distributed graph.}
\label{fig-multi}
\end{figure}

Basing on our band graphs, we have implemented a multi-sequential
refinement algorithm, outlined in Figure~\ref{fig-multi}. At every
distributed uncoarsening step, a distributed band graph is
created. Centralized copies of this band graph are then gathered on
every participating process, which serve to run fully independent
instances of our sequential FM algorithm. The perturbation of the
initial state of the sequential FM algorithm on every process allows
us to explore slightly different solution spaces, and thus to improve
refinement quality. Finally, the best refined band separator is
projected back to the distributed graph, and the uncoarsening process
goes on.

Centralizing band graphs is an acceptable solution because of the
much reduced size of the band graphs that are centralized on the
processes. Using this technique, we expect to achieve our goal,
that is, to be able partition graphs up to a billion vertices,
distributed on a thousand processes, without significant loss in
quality, because centralized band graphs will be of a size of a few
million vertices for 3D meshes.
In case the band graph cannot be centralized, we can resort to a
fully scalable algorithm, as partial copies can also be used
collectively to run a scalable parallel multi-deme genetic
optimization algorithm, such as the one experimented with
in~\cite{chpe06a}.

\section{Experimental results}
\label{secresult}

\ptscotch\ is written in ANSI C, with calls to the POSIX thread and
MPI APIs. The largest test graphs that we have used to date in our
experiments are presented in Table~\ref{tabgraphs}. All of our tests
were performed on the M3PEC system of Universit\'e Bordeaux~1, an IBM
cluster of SMP nodes made of 8 dual-core Power5 processors
running at 1.5~GHz.

\input{table_graphs.tex}

All of the ordering strategies that we have used were based on the
multi-level scheme. During the uncoarsening step, separator refinement
was performed by using our sequential FM algorithm on band graphs of
width $3$ around the projected separators, both in the parallel
(with multi-centralized copies) and in the sequential phases of the
uncoarsening process.

The quality of orderings is evaluated with respect to two
criteria. The first one, NNZ (``number of non-zeros''), is the number
of non-zero terms in the factored reordered matrix, which can be
divided by the number of non-zero terms in the initial matrix to give
the fill ratio of the ordering. The second one, OPC (``operation
count''), is the number of arithmetic operations required to factor
the matrix using the Cholesky method. It is equal to $\sum_c n_c^2$,
where $n_c$ is the number of non-zeros of column $c$ of the factored
matrix, diagonal included.

For the sake of reproducibility, the random generator used in \scotch\
is initialized with a fixed seed. This feature is essential to end
users, who can more easily reproduce their experiments and debug their
own software, and is not significant in term of performance. Indeed,
on $64$ processors, we have experimented that the maximum variation of
ordering quality, in term of OPC, between $10$ runs performed with
varying random seed, was less than $2.2$ percent on all of the above
test graphs, while running time was not impacted. Consequently,
it is not necessary for us to perform multiple runs of \scotch\ to
average quality measures.
\\

\input{table_opc_parmetis.tex}

\input{table_opc_parmetis_part2.tex}

Tables~\ref{tabparmetis} and~\ref{tabparmetisparttwo} present the OPC
computed on the orderings yielded by \ptscotch\ and \parmetis. These
results have been obtained by running \ptscotch\ with the following
default strategy: in the multi-level process, graphs are coarsened
without any folding until the average number of vertices per process
becomes smaller than $100$, after which the fold-dup process takes
place until all graphs are folded on single processes and the
sequential multi-level process relays it.

\begin{figure}[htb]
\begin{minipage}[t]{0.47\textwidth}
\includegraphics[width=\textwidth]{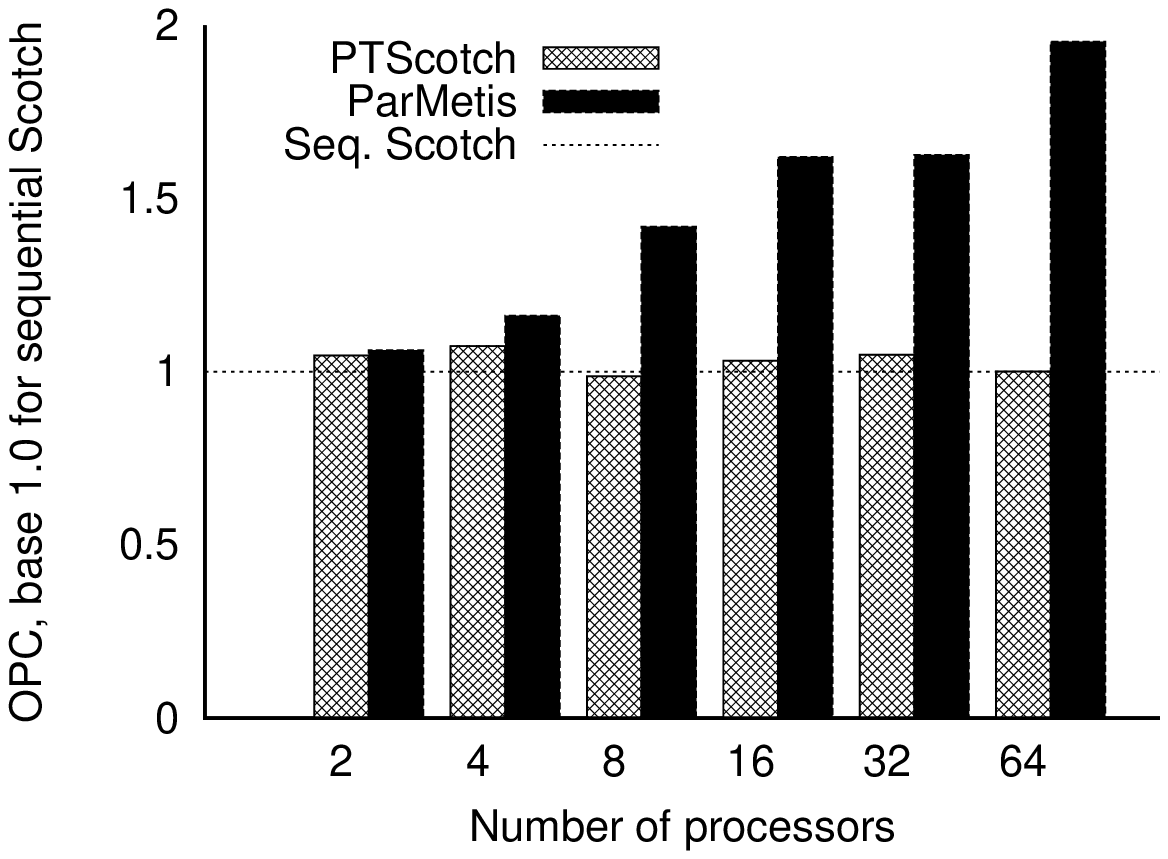}
\vspace*{-3.5em}
\caption{OPC for graph \textbf{audikw1}.}
\label{fig:audikw1-opc}
\end{minipage}%
\hspace{\fill}%
\begin{minipage}[t]{0.47\textwidth}
\includegraphics[width=\textwidth]{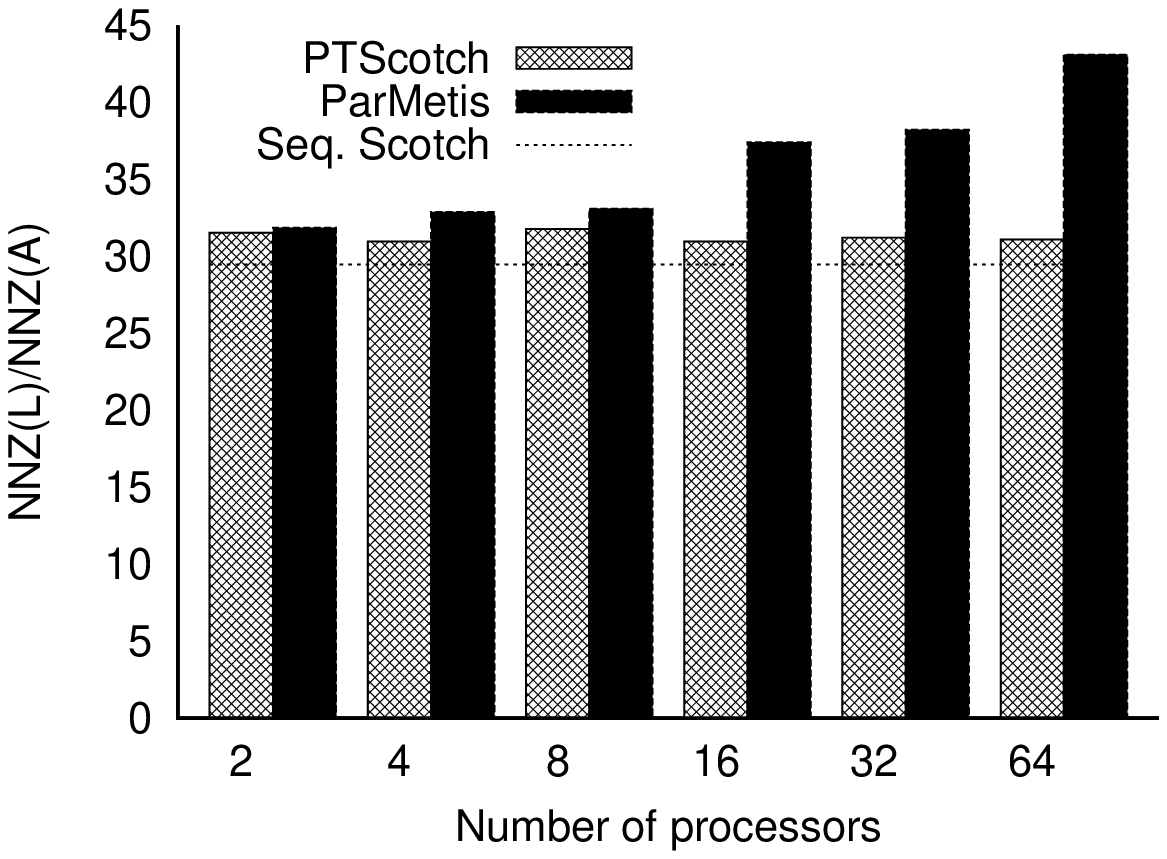}
\vspace*{-3.5em}
\caption{NNZ fill ratio for graph \textbf{audikw1}.}
\label{fig:audikw1-nnz}
\end{minipage}
\end{figure}

\begin{figure}[htb]
\begin{minipage}[t]{0.47\textwidth}
\includegraphics[width=\textwidth]{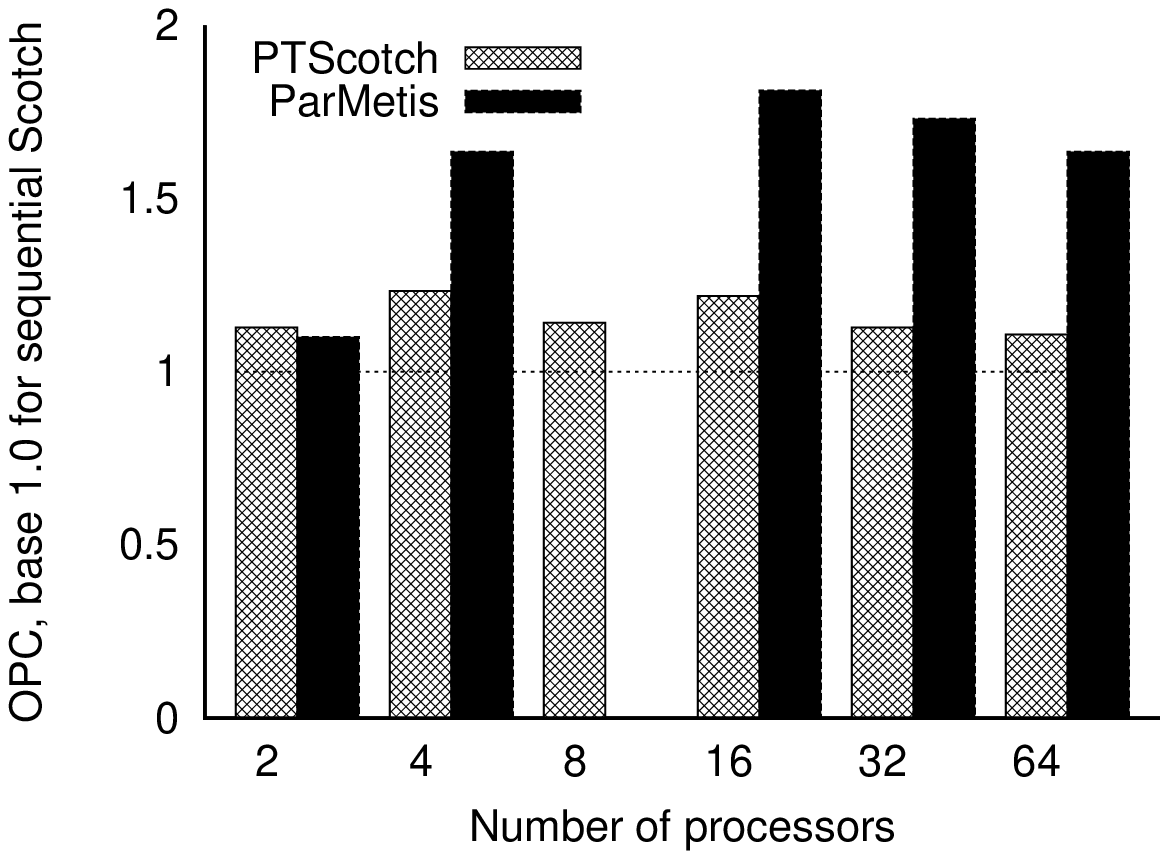}
\vspace*{-3.5em}
\caption{OPC for graph \textbf{cage15}.}
\label{fig:cage15-opc}
\end{minipage}%
\hspace{\fill}%
\begin{minipage}[t]{0.47\textwidth}
\includegraphics[width=\textwidth]{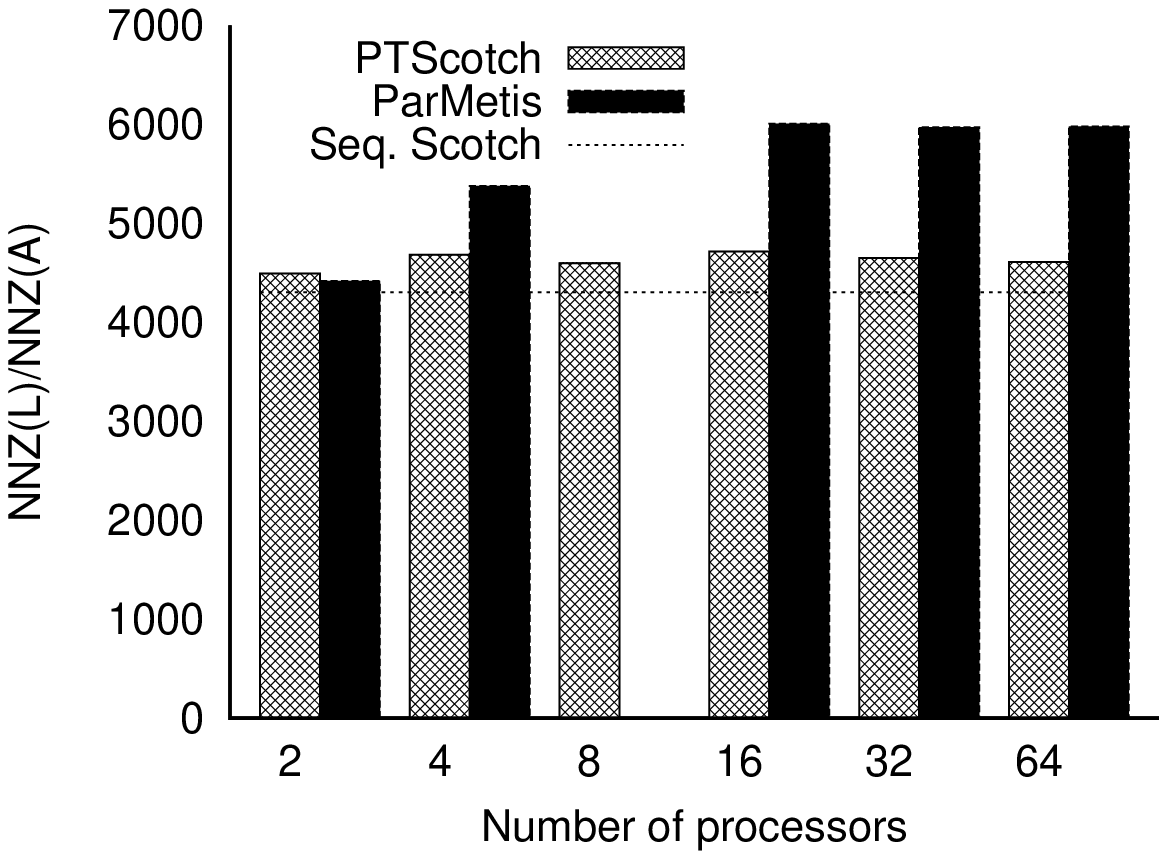}
\vspace*{-3.5em}
\caption{NNZ fill ratio for graph \textbf{cage15}.}
\label{fig:cage15-nnz}
\end{minipage}
\end{figure}

The improvement in quality brought by \ptscotch\ is clearly evidenced.
Ordering quality does not decrease along with the number of processes,
as our local optimization scheme is not sensitive to it, but instead
most often slightly increases, because of the increased number of
multi-sequential optimization steps which can be run in parallel.
Graphics in Figures~\ref{fig:audikw1-opc} to~\ref{fig:cage15-nnz}
evidence that \ptscotch\ clearly outperforms \parmetis\ in term of
ordering quality. For both graphs, the results of \ptscotch\ are
very close to the ones obtained by the sequential \scotch\ software,
while the costs of \parmetis\ orderings increase dramatically along
with the number of processes.

While \ptscotch\ is, at the time being, about four times slower on
average than \parmetis, it can yield operation counts that are as much
as two times smaller than the ones of this latter, which is of interest
as factorization times are more than one order of magnitude higher
than ordering times. For example, it takes only $41$ seconds to
\ptscotch\ to order the \textbf{brgm} matrix on $64$ processes, and
about $280$ seconds to {\sc MUMPS}~\cite{amdule00} to factorize the
reordered matrix on the same number of processes.
As the factorization process is often very scalable, it can
happen, for very large numbers of processes, that ordering times with
\ptscotch\ are higher than factorization times, because
communication latencies dominate and scalability can no longer be
guaranteed. Because of their different complexity and scalability
behaviors, it is in general not useful to run the ordering tool on as
many processes as the solver; what matters is to have enough
distributed memory to store the graph and run the ordering process,
while keeping good scalability in time.
Moreover, in various applications, the ordering phase is performed
only once while the factorization step is repeated many times with
different numerical values, the matrix structure being invariant. This
is the reason why we essentially focused on ordering quality rather
than on optimal scalability.

The multi-sequential FM refinement algorithm that we use is by nature
not scalable, and its cost, even on band graphs, is bound to dominate
parallel execution time when the number of processors increases. This
is why on some graphs, depending on their size, average degree and
topology, running times of \ptscotch\ no longer decrease. We are
therefore still working on improving the scalability in time of
\ptscotch, and especially of its coarsening phase, which is the most
time-consuming algorithm in term of communications. We are also
investigating several ways to replace the intrinsically sequential FM
refinement algorithm by fully parallel algorithms such as parallel
diffusion-based methods~\cite{pell07b} or genetic
algorithms~\cite{chpe06a}.

\begin{figure}[htb]
\begin{minipage}[t]{0.47\textwidth}
\includegraphics[width=\textwidth]{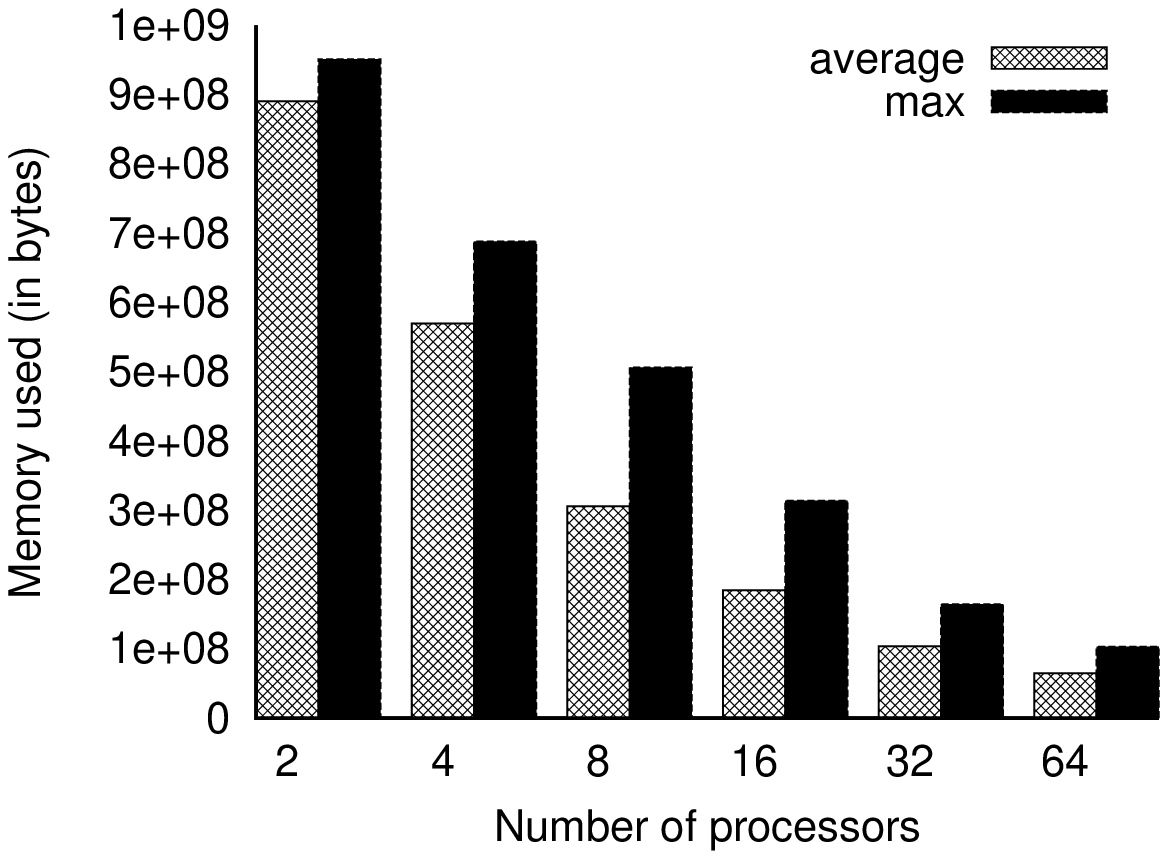}
\vspace*{-3.5em}
\caption{Memory used per process to reorder graph \textbf{audikw1}.}
\label{fig:audikw1-mem}
\end{minipage}%
\hspace{\fill}%
\begin{minipage}[t]{0.47\textwidth}
\includegraphics[width=\textwidth]{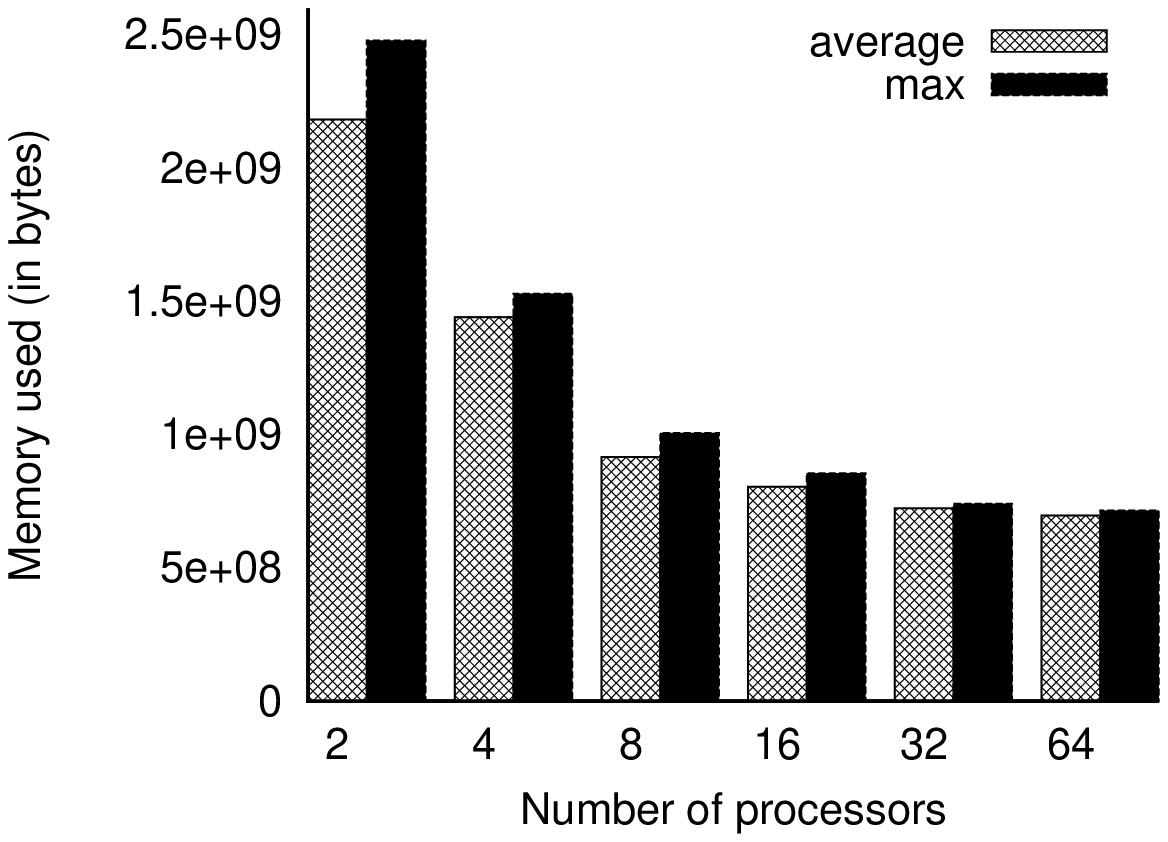}
\vspace*{-3.5em}
\caption{Memory used per process to reorder graph \textbf{cage15}.}
\label{fig:cage15-mem}
\end{minipage}
\end{figure}

Figures~\ref{fig:audikw1-mem} and~\ref{fig:cage15-mem} show the amount
of memory used per process during the ordering of two of our test
graphs. Figure~\ref{fig:audikw1-mem} evidences that, despite the use
of folding and duplication at the coarsest levels, memory scalability
remains good. However, imbalance can be high, especially for graph
\textbf{audikw1}. Because this graph possesses a set of contiguous
vertices of very high degree, and as all of our data distributions
balance numbers of vertices only, there can be cases in which one
process holds most of the edges by owning entirely the subset of
highly connected vertices. Also, for graph \textbf{cage15}, memory
scalability can only be observed between two and eight processes. The
main reason is that the vertices of this graph are initially ordered
such that the number of ghost vertices quickly increases when the
number of processes reaches $16$. However, since this graph is of low
degree, memory scalability will eventually be evidenced again, in term
of edges rather than of vertices, after all edges are considered to be
connected to ghost vertices.


\section{Conclusion}
\label{secconcl}

We have presented in this paper the parallel algorithms that we have
implemented in \ptscotch\ to compute in parallel efficient orderings
of very large graphs. The first results are encouraging, as they meet
the expected performance requirements in term of quality, but have yet
to be improved in term of scalability. Indeed, scalability can only be
partially evidenced for very large graphs, because of the lack of
scalability of our multi-sequential band refinement algorithm.
However, we have been able to run test graphs which could not fit in
memory when processed by competing software.

To date, we have been able to order 3D graphs up to $23$ million
vertices, which have been successfully factorized afterwards, in double
precision complex arithmetic, by the \pastix\ parallel direct
solver~\cite{heraro05} also developed within the \scalapplix\ project.
Although existing parallel direct sparse linear system
solvers cannot currently handle full 3D meshes of sizes larger than
about fifty million unknowns, we plan to use \ptscotch\ to compute
orderings of larger matrices to be factorized using hybrid
direct-iterative schemes~\cite{hesa06,heraro07}.
\\

However, sparse matrix ordering is not the application field in which
we expect to find the largest problem graphs. Basing on the software
building blocks that we have already written, we plan to extend the
capabilities of \ptscotch\ to compute edge-separated k-ary partitions
of large meshes for subdomain-based fully iterative methods, as well
as static mappings of process graphs, like the \scotch\ library does
sequentially.




\end{document}

%% file: table_graphs.tex

\begin{table}[Hbt]
\begin{center}
  \begin{tabular}{|l||r|r|r|c|r|}
    \hline
    \multicolumn{1}{|c||}{\smash{\raisebox{-0.5em}{\bf Graph}}} &
    \multicolumn{2}{c|}{{\bf Size} ($\times 10^{3}$)} &
    \multicolumn{1}{c|}{\bf Average} &
    \multicolumn{1}{|c|}{\smash{\raisebox{-0.5em}{$O_{SS}$}}} &
    \multicolumn{1}{|c|}{\smash{\raisebox{-0.5em}{\bf Description}}}\\
    \cline{2-3}
    ~ & \multicolumn{1}{c|}{$|V|$} & \multicolumn{1}{c|}{$|E|$} &
    \multicolumn{1}{c|}{\bf degree} & ~ & ~\\
    \hline
    \hline
    23millions & $23114$ & $175686$ & $7.60$ & 1.29e+14 & 3D electromagnetics, CEA \\
    altr4 & $26$ & $163$ & $12.50$ & 3.65e+8 & 3D electromagnetics, CEA \\
    audikw1 & $944$ & $38354$ & $81.28$ & 5.48e+12 & 3D mechanics mesh, Parasol \\
    bmw32 & $227$ & $5531$ & $48.65$ & 2.80e+10 & 3D mechanics mesh, Parasol \\
    brgm & $3699$ & $151940$ & $82.14$ & 2.70e+13 & 3D geophysics mesh, BRGM \\
    cage15 & $5154$ & $47022$& $18.24$ & 4.06e+16 & DNA electrophoresis, UF \\
    conesphere1m & $1055$ & $8023$ & $15.21$ & 1.83e+12 & 3D electromagnetics, CEA \\
    coupole8000 & $1768$ & $41657$ & $47.12$ & 7.46e+10 & 3D structural mechanics, CEA \\
    qimonda07 & $8613$ & $29143$ & $6.76$ & 8.92e+10 & Circuit simulation, Qimonda \\
    thread & $30$ & $2220$ & $149.32$ & 4.14e+10 & Connector problem, Parasol \\
    \hline
  \end{tabular}
\end{center}
\caption{Description of some of the test graphs that we use. $|V|$ and
  $|E|$ are the vertex and edge cardinalities, in thousands, and $O_{SS}$ is
  the operation count of the Cholesky factorization performed on
  orderings computed using the sequential \scotch\ software. CEA is
  the French atomic energy agency, UF stands
  for the University of Florida sparse matrix
  collection~\cite{florida}, and Parasol is a former European
  project~\cite{parasol}.}
\label{tabgraphs}
\end{table}

%% file: table_opc_parmetis.tex

\begin{table}[hbt]
\begin{center}
  \begin{tabular}{|l||c|c|c|c|c|c|}
    \hline
    \multicolumn{1}{|c||}{{\bf Test}} &
    \multicolumn{6}{c|}{{\bf Number of processes}}\\
    \cline{2-7}
     \multicolumn{1}{|c||}{{\bf case}} & 2 & 4 & 8 & 16 & 32 & 64 \\
    \hline
    \hline
    \multicolumn{7}{|c|}{{\bf 23millions}} \\
    \hline
    $O_{PTS}$ & \textbf{1.45e+14} & \textbf{2.91e+14} & \textbf{3.99e+14} & \textbf{2.71e+14} & \textbf{1.94e+14} & \textbf{2.45e+14} \\
    $O_{PM}$  & $\dag$  &$\dag$ & $\dag$ & $\dag$ & $\dag$ & $\dag$   \\
    $t_{PTS}$ & 671.60 & 416.75 & 295.38 & 211.68 & 147.35 & 103.73 \\
    $t_{PM}$  & $\dag$  &$\dag$ & $\dag$ & $\dag$ & $\dag$ & $\dag$ \\
    \hline
    \multicolumn{7}{|c|}{{\bf altr4}} \\
    \hline
    $O_{PTS}$ & \textbf{3.84e+8}  & \textbf{3.75e+8}  & \textbf{3.93e+8}  & \textbf{3.69e+8}  & \textbf{4.09e+8}  & \textbf{4.15e+8}   \\
    $O_{PM}$  & 4.20e+8  & 4.49e+8  & 4.46e+8  & 4.64e+8  & 5.03e+8  & 5.16e+8  \\
    $t_{PTS}$ & 0.42     & 0.30     & 0.24     & 0.30     & 0.52     & 1.55     \\
    $t_{PM}$  & 0.31     & 0.20     & 0.13     & 0.11     & 0.13     & 0.33	\\
    \hline
    \multicolumn{7}{|c|}{{\bf audikw1}} \\
    \hline
    $O_{PTS}$ & \textbf{5.73e+12} & \textbf{5.65e+12} & \textbf{5.54e+12} & \textbf{5.45e+12} & \textbf{5.45e+12} & \textbf{5.45e+12} \\
    $O_{PM}$  &  5.82e+12  & 6.37e+12 & 7.78e+12 & 8.88e+12 & 8.91e+12 & 1.07e+13 \\
    $t_{PTS}$ & 73.11    & 53.19   & 45.19    & 33.83   &  24.74    & 18.16   \\
    $t_{PM}$  & 32.69    & 23.09    & 17.15    & 9.804    & 5.65     & 3.82     \\
    \hline
    \multicolumn{7}{|c|}{{\bf bmw32}} \\
    \hline
    $O_{PTS}$ & 3.50e+10 & \textbf{3.49e+10} & \textbf{3.14e+10} & \textbf{3.05e+10}  & \textbf{3.02e+10} & \textbf{3.00e+10} \\
    $O_{PM}$  & \textbf{3.22e+10} & 4.09e+10 & 5.11e+10 & 5.61e+10  & 5.74e+10  & 6.31e+10  \\
    $t_{PTS}$ & 8.89     & 7.41     & 5.68     & 5.45     & 8.36     & 17.64   \\
    $t_{PM}$  & 3.39     & 2.28	    & 1.51     & 0.92     & 0.68     & 1.08     \\
    \hline
    \multicolumn{7}{|c|}{{\bf brgm}} \\
    \hline
    $O_{PTS}$ & \textbf{2.70e+13} & \textbf{2.55e+13} & \textbf{2.65e+13} & \textbf{2.88e+13} & \textbf{2.86e+13} & \textbf{2.87e+13} \\
    $O_{PM}$  & -- & $\dag$ & $\dag$ & $\dag$ & $\dag$ & $\dag$ \\
    $t_{PTS}$ & 276.9 & 167.26 & 97.69 & 61.65 & 42.85 & 41.00 \\
    $t_{PM}$  & -- & $\dag$ & $\dag$ & $\dag$ & $\dag$ & $\dag$ \\
    \hline
    \multicolumn{7}{|c|}{{\bf cage15}} \\
    \hline
    $O_{PTS}$ & 4.58e+16 & \textbf{5.01e+16} & \textbf{4.64e+16} & \textbf{4.95e+16} & \textbf{4.58e+16} & \textbf{4.50e+16} \\
    $O_{PM}$  & \textbf{4.47e+16} & 6.64e+16 & $\dag$ & 7.36e+16 & 7.03e+16 & 6.64e+16\\
    $t_{PTS}$ & 540.46 & 427.38 & 371.70 & 340.78 & 351.38 & 380.69 \\
    $t_{PM}$  & 195.93 & 117.77 & $\dag$ & 40.30 & 22.56 & 17.83 \\
    \hline
    \multicolumn{7}{|c|}{{\bf conesphere1m}} \\
    \hline
    $O_{PTS}$ & \textbf{1.88e+12} & \textbf{1.89e+12} & \textbf{1.85e+12} & \textbf{1.84e+12} & \textbf{1.86e+12} & \textbf{1.77e+12} \\
    $O_{PM}$  & 2.20e+12 & 2.46e+12 & 2.78e+12 & 2.96e+12 & 2.99e+12 & 3.29e+12 \\
    $t_{PTS}$ & 31.34    & 20.41    & 18.76    & 18.37    & 25.80    & 92.47    \\
    $t_{PM}$  & 22.40    & 11.98    & 6.75     & 3.89     & 2.28     & 1.87     \\
    \hline
  \end{tabular}
\end{center}
\caption{Comparison between \ptscotch\ (PTS) and \parmetis\ (PM) for
  several graphs. $O_{PTS}$ and $O_{PM}$ are the OPC for PTS and PM,
  respectively. Dashes indicate abortion due to memory shortage.
  Daggers indicate abortion due to an invalid MPI operation.}
\label{tabparmetis}
\end{table}

%% file: table_opc_parmetis_part2.tex

\begin{table}[hbt]
\begin{center}
  \begin{tabular}{|l||c|c|c|c|c|c|}
    \hline
    \multicolumn{1}{|c||}{{\bf Test}} &
    \multicolumn{6}{c|}{{\bf Number of processes}}\\
    \cline{2-7}
     \multicolumn{1}{|c||}{{\bf case}} & 2 & 4 & 8 & 16 & 32 & 64 \\
    \hline
    \hline
    \multicolumn{7}{|c|}{{\bf coupole8000}} \\
    \hline
    $O_{PTS}$ & \textbf{8.68e+10} & \textbf{8.54e+10} & 8.38e+10 & \textbf{8.03e+10} & \textbf{8.26e+10} & \textbf{8.21e+10} \\
    $O_{PM}$  & $\dag$   & $\dag$   & \textbf{8.17e+10} & 8.26e+10 & 8.58e+10 & 8.71e+10 \\
    $t_{PTS}$ & 114.41   & 116.83   & 85.80    & 60.23    & 41.60    & 28.10    \\
    $t_{PM}$  & 63.44    & 37.50    & 20.01    & 10.81    & 5.88     & 3.14     \\
    \hline
    \multicolumn{7}{|c|}{{\bf qimonda07}} \\
    \hline
    $O_{PTS}$ & -- & -- & \textbf{5.80e+10} & \textbf{6.38e+10} & \textbf{6.94e+10} & \textbf{7.70e+10} \\
    $O_{PM}$  & $\dag$ & $\dag$ & $\dag$ & $\dag$ & $\dag$   & $\dag$  \\
    $t_{PTS}$ & --  & -- & 34.68 & 22.23 & 17.30 & 16.62 \\
    $t_{PM}$  & $\dag$  &$\dag$ & $\dag$ & $\dag$ & $\dag$  & $\dag$  \\
    \hline
    \multicolumn{7}{|c|}{{\bf thread}} \\
    \hline
    $O_{PTS}$ & \textbf{3.52e+10} & \textbf{4.31e+10} & \textbf{4.13e+10} & \textbf{4.06e+10} & \textbf{4.06e+10} & \textbf{4.50e+10} \\
    $O_{PM}$  & 3.98e+10 & 6.60e+10 & 1.03e+11 & 1.24e+11 & 1.53e+11 & --       \\
    $t_{PTS}$ & 3.66     & 3.61     & 3.30     & 3.65     & 5.68     & 11.16    \\
    $t_{PM}$  & 1.25     & 1.05     & 0.68     &  0.51    & 0.40     & --       \\
    \hline
  \end{tabular}
\end{center}
\caption{Continuation of Table~\protect{\ref{tabparmetis}}.}
\label{tabparmetisparttwo}
\end{table}